\documentclass[conference,a4paper]{IEEEtran}
\usepackage{graphicx}
\usepackage{multirow}
\usepackage{multicol}
\usepackage[utf8]{inputenc}
\usepackage[english]{babel}
\usepackage{rotating, graphicx}
\usepackage{array,multirow,graphicx}
\usepackage{amsmath}
\usepackage{amsfonts}
\usepackage{lipsum}
\usepackage{mathtools}
\usepackage{float}
\usepackage[caption = false]{subfig}
\usepackage{amsmath} 
\usepackage{hyperref}
\usepackage{fancyhdr}
\usepackage{booktabs,caption}
\usepackage{threeparttable}
\begin{document}

\title{How Should Network Slice Instances be Provided to Multiple Use Cases of a Single Vertical Industry?}

\author{\authorblockN{Mohammad Asif Habibi\IEEEauthorrefmark{1}, Bin Han\IEEEauthorrefmark{1}, Faqir Zarrar Yousaf\IEEEauthorrefmark{2}, and Hans D. Schotten\IEEEauthorrefmark{1}\IEEEauthorrefmark{3}\\ \IEEEauthorblockA{\IEEEauthorrefmark{1}Technische Universit\"at Kaiserslautern, \IEEEauthorrefmark{2}NEC Laboratories Europe, \IEEEauthorrefmark{3}German Research Center for Artificial Intelligence }}}

\maketitle

\begin{abstract}
There are a large number of vertical industries implementing multiple use cases, each use case characterized by diverging service, network, and connectivity requirements -- such as automobile, manufacturing, power grid, etc. Such heterogeneity cannot be effectively managed and efficiently mapped onto a single type of network slice instance (NSI). Thus the tailored provisioning of an end-to-end network slicing solution to a vertical industry that consists of multiple use cases is a critical issue, which motivates this article, aimed at exploring this never-addressed and challenging research problem by proposing the Use-case Specific Network Slicing and the Sub Network Slicing concepts that enable the provisioning of Use-case Specific NSI (US-NSI) and GeNeric NSI (GN-NSI), respectively. Both approaches tackle the same technical issue of provisioning, management, and orchestration of per–vertical per–use-case NSIs in order to improve resource allocation and enhance network performance. The article also presents the architectural frameworks for managing US-NSI and GN-NSI, which extend the service deployment concept and system architecture of network slicing for vertical customers of fifth generation (5G) mobile networks.
\end{abstract}
\IEEEoverridecommandlockouts
\IEEEpeerreviewmaketitle

\section{Introduction}\label{sec:Introduction}
With the deployment of network slicing in the fifth generation (5G) system, the communication service provider (hereinafter operator) is able to partition a physical network into a number of virtual (logical) networks -- each referred to as a network slice instance (NSI) -- through network function virtualization (NFV) and software-defined networking (SDN) technologies \cite{7926923}. An NSI is instantiated based on a template (sometimes also referred to as blueprint, service profile, and descriptor), which is a set of attributes with values/ranges, known as the network slice template (NST). The NST defines network resources/functions, and their interconnections and configurations to meet the diverse technical requirements of a use case \cite{NGMNConcept}. The NSI is offered by an operator to a communication service customer (hereinafter tenant) as a service \cite{7509393}. Each tenant requires a formal service level agreement (SLA) with the operator in order to assess the characteristics of an NSI, and the responsibilities and priorities of both parties. 

\begin{figure*}[h]
    \centering
    \includegraphics[scale=0.5]{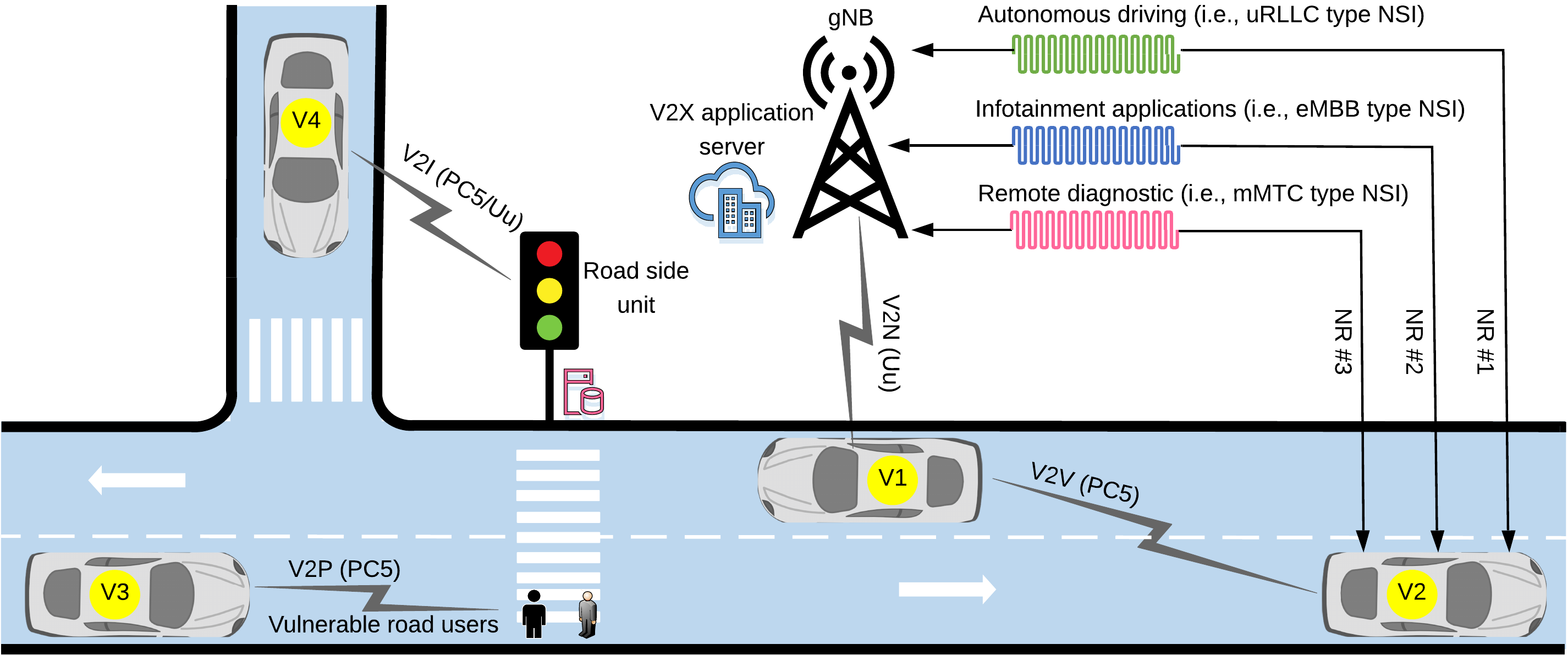}
    \caption{Three use cases and their corresponding NSIs of the V2X communication system. Each NSI is individually deployed with configurable characteristics and different levels of isolation to only one use case out of a group of use cases.}
    \label{fig:concept.pdf}
\end{figure*}

To date, a large number of theoretical researches, industrial experiments, and standardization activities have been carried out with the purpose of exploring various dimensions of network slicing (such as resource allocation, admission control, performance optimization, management, orchestration, etc.) in the context of a single NSI. Usually, an NSI is configured to support multiple service instances belonging to the same service category: enhanced mobile broadband (eMBB), massive machine-type communication (mMTC), or ultra-reliable low latency communication (uRLLC) for a particular vertical industry (hereinafter vertical). However, such service provisioning is not optimal as each service instance within the same vertical may represent a different use case with diverging quality of service (QoS), configuration, and resource requirements \cite{7926923} \cite{NGMNConcept}. For example, the power grid industry contains a group of use cases including distribution automation, net metering, renewable power supply, etc. Every use case (out of a family of use cases of a vertical) has its own specific performance, functional, and operational requirements. Therefore, one NSI per single tenant is not able to fulfill their diverse service demands.

This issue is partially raised by Global System for Mobile communications Association (GSMA) in \cite{GSMA}, International Telecommunication Union (ITU) in \cite{ITU-G.7702}, and Open Networking Foundation (ONF) in \cite{ONFSDNArchitecture}. The European Telecommunications Standards Institute (ETSI) NFV, Third Generation Partnership Project (3GPP), Next Generation Mobile Networks (NGMN) Alliance, and Internet Engineering Task Force (IETF) have not provided solutions for supporting divergent use cases of a vertical in their standardization activities so far. Providing network slicing solutions -- mainly the management and orchestration of NSIs, and allocating their corresponding physical and virtual resources -- to such type of verticals is still an open issue. Addressing them, in the standards development organizations (SDOs) especially in 3GPP and ETSI NFV, are thus essential to design a complete interoperable end-to-end (E2E) network slicing paradigm.

Motivated by the solution gap described above, we propose two concepts, namely the Use-case Specific Network Slicing and the Sub Network Slicing, and their architectural frameworks which enable the provisioning of Use-case Specific NSI (US-NSI) and the GeNeric NSI (GN-NSI). The objective is to provide solution proposals on this crucial issue of network slicing in SDOs (see Sec. \ref{sec:Use-caseSpecific} and Sec. \ref{sec:GenericNetworkSlice}, respectively). Both approaches are in compliance with the E2E service deployment concept and their architectures are built on top of architectural frameworks of network slicing, defined by the NGMN Alliance in \cite{NGMNConcept} and standardized by the 3GPP in \cite{3GPPTS23501V1540}. The main objectives of the proposed solutions are to provide an insight into the concepts, use cases and their integration in 3GPP networks, procedures, and mechanisms required to make the deployment of independent NSIs to each of the use cases of such verticals in a cost/resource-efficient, flexible, and agile manner from the perspective of both operator and tenant. Before addressing these issues, we commence by providing a background on key concepts in subsequent section \ref{sec:Background}. 

\section{Network Slicing for Multiple Use Cases of a Single Vertical} \label{sec:Background}
Every vertical is owned by at least a single tenant and consists of at least one use case. Every use case is represented by at least one service instance and thus requires its own architected and optimized NSI with a customized NST. Each NSI dedicated to a specific use case is isolated in terms of resources, security, policies, configuration, etc. from other NSIs and may contain specific capabilities in 5G core network (5GC), next-generation radio access network (NG-RAN), transport network (TN), and terminal \cite{Ferrus2019}. Each NSI is managed by its own manager and has operation and maintenance (O\&M) requirements unique to the NSI. The O\&M system is responsible to provide monitoring and self-control capabilities to support business requirements throughout the life-cycle of an NSI. Therefore, the interactions between tenant and operator during the business period is constructively handled in order for both parties to negotiate their conflicting interests and to achieve satisfaction.

The design, the negotiation of SLA, and the life-cycle management (LCM) of an NSI dedicated to a vertical that consists of a single use case are straightforward processes. The technical requirements of such verticals can be mapped onto the E2E network slicing concept, easily and efficiently. However, verticals that contain more than one use case are usually characterized by diverse service requirements. For example, a vehicle to everything (V2X) system consists of various use cases in four communication modes (see Fig. \ref{fig:concept.pdf}) -- vehicle to vehicle (V2V), vehicle to infrastructure (V2I), vehicle to pedestrian (V2P), and vehicle to network (V2N) -- defined by the 3GPP \cite{3GPPTR23785V110}. Among them, autonomous driving, infotainment applications, and remote diagnostic are the most well-known use cases. Every use case demands its own set of QoS requirements tailored to satisfy specific business requirements of one or more V2X services (see Tab. \ref{Tab:UseCaseRequirements}). Therefore, each use case requires an individual and isolated NSI. These NSIs are intended for different scenarios. For instance, uRLLC for autonomous driving, eMBB for infotainment applications, and mMTC for remote diagnostics.

\newcommand{\STAB}[1]{\begin{tabular}{@{}c@{}}#1\end{tabular}}
\begin{table*}[ht]
\centering
\caption{Quantitative and qualitative comparison of performance, functional, and operational requirements of three NSIs of the V2X system.}
\begin{tabular}{|c|l|l|l|l|}
\hline \hline
\multirow{2}{*}{\textbf{Category}} & \multicolumn{1}{c|}{\multirow{2}{*}{\textbf{Requirement}}} & \multicolumn{3}{c|}{\textbf{Three NSIs supporting three use cases of the V2X system}}                                                                                                                                                                                          
\\ \cline{3-5} 
                                      & \multicolumn{1}{c|}{}                                      & \multicolumn{1}{c|}{\textbf{\begin{tabular}[c]{@{}c@{}}uRLLC type NSI \\ for autonomous driving \\ (V2I, V2N, V2V)\end{tabular}}} & \multicolumn{1}{c|}{\textbf{\begin{tabular}[c]{@{}c@{}}eMBB type NSI \\ for infotainment \\ (V2N)\end{tabular}}} & \multicolumn{1}{c|}{\textbf{\begin{tabular}[c]{@{}c@{}}mMTC type NSI \\ for remote  diagnostics \\ (V2I, V2N)\end{tabular}}} \\ \hline
\multirow{6}{*}{\STAB{\rotatebox[origin=c]{90}{\textbf{Performance}}}}  & Latency                                                    & 1-10 ms                                                                                                                                     & \textless{}20 ms                                                                                                           & \textless{}100 ms                                                                                                                      \\ \cline{2-5} 
                                      & Reliability                                                & 99.999\%                                                                                                                                    & 99.99\%                                                                                                                    & 95\%                                                                                                                                   \\ \cline{2-5} 
                                      & Availability                                               & 99.9999\%                                                                                                                                   & 99.999\%                                                                                                                   & 99\%                                                                                                                                   \\ \cline{2-5} 
                                      & Mobility                                                   & 0-250 Km/hr                                                                                                                                 & 0-250 Km/hr                                                                                                                & 0-250 Km/hr                                                                                                                            \\ \cline{2-5} 
                                      & Device density                                             & High                                                                                                                                        & High                                                                                                                       & Very high                                                                                                                              \\ \cline{2-5} 
                                      & Data rate                                                  & 50 Mbps                                                                                                                                     & 1-100 Mbps                                                                                                                 & 0.55 Mbps                                                                                                                              \\ \hline
\multirow{6}{*}{\STAB{\rotatebox[origin=c]{90}{\textbf{Functional}}}} & Isolation                                                  & Very high                                                                                                                                   & High                                                                                                                       & Medium                                                                                                                                 \\ \cline{2-5} 
                                      & Security                                                   & Very high                                                                                                                                   & High                                                                                                                       & Not a concern                                                                                                                          \\ \cline{2-5} 
                                      & Application server positioning                                             & Not required                                                                                                                                & Edge/remote-cloud                                                                                                          & Remote cloud                                                                                                                           \\ \cline{2-5} 
                                      & Scheduling                                           & Semi-persistent                                                                                                                             & Dynamic                                                                                                                    & Semi-persistent                                                                                                                        \\ \cline{2-5} 
                                      & Priority                                                   & Very high                                                                                                                                   & High                                                                                                                       & Medium                                                                                                                                 \\ \cline{2-5} 
                                      & Battery life                                               & High                                                                                                                                        & High                                                                                                                       & Very high                                                                                                                              \\ \hline
\multirow{6}{*}{\STAB{\rotatebox[origin=c]{90}{\textbf{Operational}}}}  & Coverage type                                              & Nationwide                                                                                                                                  & Global                                                                                                                     & Nationwide                                                                                                                             \\ \cline{2-5} 
                                      & Supported APIs                                             & Yes                                                                                                                                         & Yes                                                                                                                        & Yes                                                                                                                                    \\ \cline{2-5} 
                                                                            & Energy efficiency                                          & High                                                                                                                                        & High                                                                                                                       & Very high                                                                                                                              \\ \cline{2-5} 
                                      & Resources/policies                                                 & Self management                                                                                                                                 & Self management                                                                                                                & Self management                                                                                                                            \\ \cline{2-5} 
                                      & Monitoring                                                 & Real                                                                                                                                        & Real/non-real                                                                                                              & Real/non-real                                                                                                                          \\ \cline{2-5} 
                                      & Communication mode                                                 & PC5                                                                                                                              & LTE-Uu/NR                                                                                                             & LTE-Uu/NR                                                                                                                         \\ \cline{2-5} 
                                      & Communication primitive                                            & Broadcast                                                                                                                                   & Unicast                                                                                                                    & Unicast                                                                                                                                \\ \hline
\end{tabular}
\begin{tablenotes}
      \small
      \item \textbf{Note:} Performance requirements are the most demanding requirements, which specify the characteristics and/or type of an NSI. Functional requirements are some of the basic functions that are used to describe the intended capabilities and interactions of an NSI. Operational requirements are associated with the operation, mission, and management of an NSI. Please also note that all values in Tab. \ref{Tab:UseCaseRequirements} are in compliance with 3GPP Releases 14 -- 16.
    \end{tablenotes}
\label{Tab:UseCaseRequirements}
\end{table*}

\begin{itemize}
\item NSI supporting autonomous driving requires ultra-high reliability (99.999\%) and ultra-low latency (below 10 ms) on V2V mode over sidelink air-interface (PC5). Moreover, full-road network coverage (99.9999\%) is needed to provide services for autonomous vehicles in all types of  geographical locations under high vehicle-density condition (see Tab. \ref{Tab:UseCaseRequirements}). This NSI, in some scenarios, may also provide driving efficiency and safety services, such as using a see-through application to detect hazards on road and to trigger emergency messages.

\item NSI for infotainment applications must be configured to offer a short-lived high data rate connection of up to 100 Mbps (for mobile broadband in vehicles) between a vehicle and nearby next-generation NodeBs (gNBs) in order to support services, such as the broadband transmission of video, files/apps downloading/uploading, online gaming, access to social media, audio/video conference streaming, etc. This NSI should support connection under mobility of up to 250 Km/hr on V2I and V2N modes over the air interface (see Tab. \ref{Tab:UseCaseRequirements}).

\item NSI designed for vehicle management and remote diagnostic should provide connectivity between a massive number of vehicles (hundreds per kilometer square, see Tab. \ref{Tab:UseCaseRequirements}) and a remote server outside a communication network. This NSI is functioning on V2I and V2N modes over long term evolution (LTE)-Uu/5G new radio (NR) air interfaces and meanwhile demands extreme power saving mechanism. It is usually owned by a V2X service provider or diagnostic center that continuously collects information from sensors installed on a vehicle to track the position of the vehicle, the status of the vehicle, perform remote diagnostic troubleshooting, etc.
\end{itemize}

The example of instantiating three E2E isolated NSIs to three use cases of the V2X system proves that providing network slicing solutions to verticals that consist of multiple use cases is an important practical research problem. Identifying the existing use cases and fulfilling the QoS requirements of such verticals by providing per--use-case NSIs brings a number of key advantages to both tenant and operator. From the tenant’s perspective, the proposed approach simplifies the operations by enforcing the QoS requirements of individual use cases in isolation from others. In this way, the tenant’s quality of experience (QoE) is improved and the business flexibility is also empowered. Whereas, from the operator’s perspective, both physical and virtual resources are efficiently allocated to the individual use case of a vertical.  

\begin{figure*}[!h]
    \centering
    \includegraphics[scale=0.5]{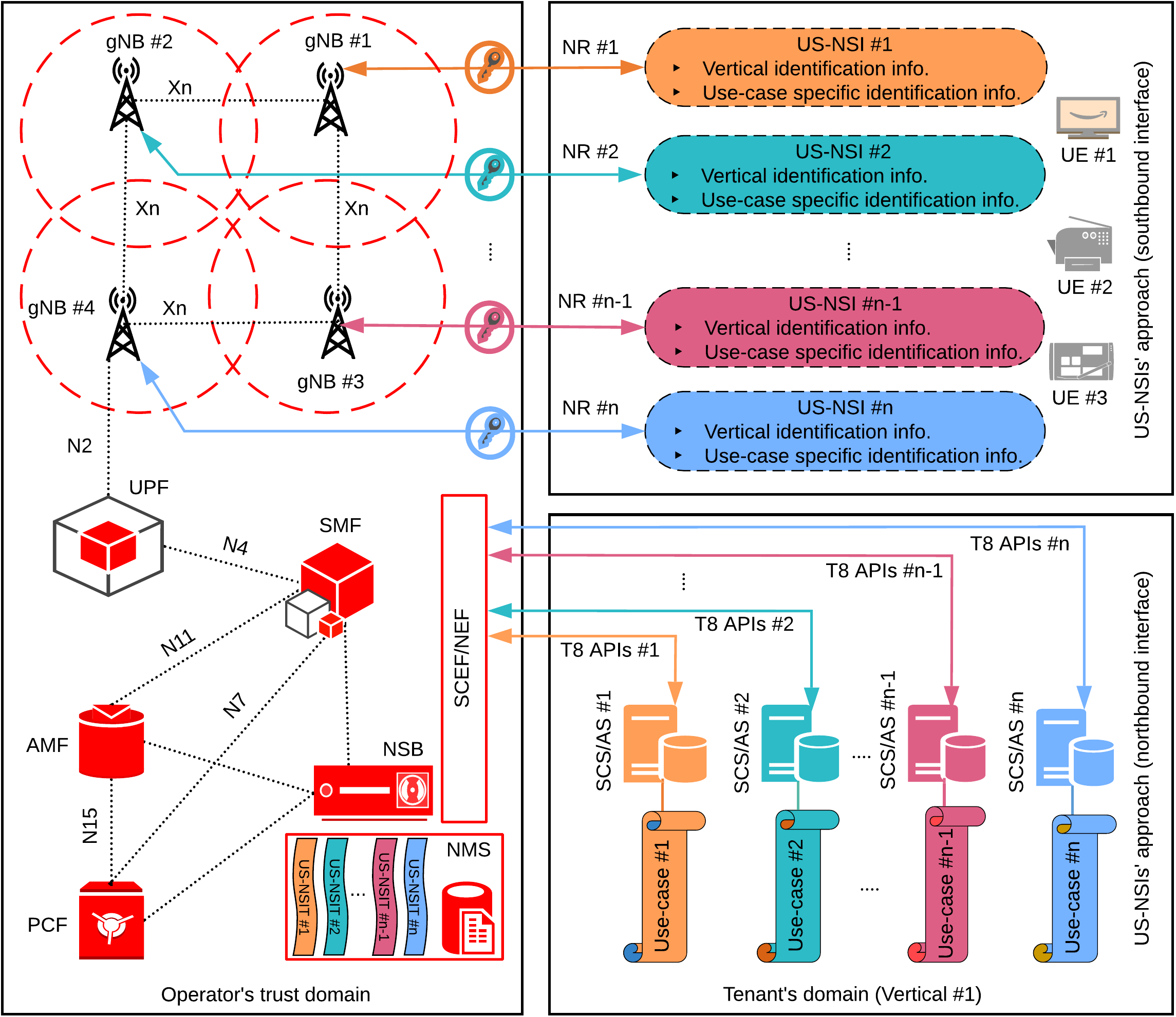}
    \caption{The architectural framework of the Use-case Specific Network Slicing.}
    \label{fig:US-NSI.pdf}
\end{figure*}

The per-use-case NSIs can be dynamically offered by either a single or multiple operators. In this article, a single operator is assumed. To offer these NSIs, either of the following proposed approaches -- the Use-case Specific Network Slicing or the Sub Network Slicing -- should be chosen. 

\begin{itemize}
\item E2E Use-case Specific Network Slicing: in this approach, a per-use-case NSI (known as the US-NSI) is provisioned, where the number of the US-NSIs equals the number of use cases.
\item E2E Sub Network Slicing: in this approach, a per-vertical GN-NSI is provisioned, which contains a family of Sub Network Slice Instances (S-NSIs), where every S-NSI is dedicated to a use case.
\end{itemize}

In both approaches, all existing use cases have to be identified first. In order to ensure specific needs and to meet diverse service demands of every use case, all of its performance, functional, and operational attributes should be then individually defined and quantified. Subsequently, the NST of every use case shall be designed and uploaded into the network slice catalogue (NSC).

\section{Use-case Specific Network Slicing} \label{sec:Use-caseSpecific}
In this concept, a per-use-case NSI, namely the US-NSI, is instantiated to support a specific use case (out of a group of use cases) of a vertical. The US-NSIs are provided according to the number of the use cases of a vertical. For example, if there are three use cases in the V2X communication system, then three US-NSIs must be offered. The LCM of a US-NSI is isolated from other US-NSIs of the same vertical. In other words, every US-NSI is individually activated, operated, and deactivated. Therefore, every US-NSI must also have its own SLA.

\subsection{The Architectural Framework of the Use-case Specific Network Slicing}
In Fig. \ref{fig:US-NSI.pdf}, Vertical \#1 owned by Tenant \#1 that consist of $n\in\mathbb{N}^+$ use cases (i.e., Use-case \#1 -- Use-case \#n) is assumed. Tenant \#1 requests $n$ US-NSIs (i.e., US-NSI \#1 -- US-NSI \#n, respectively). The number of slice requests in the northbound interface thus equals the number of use cases of Vertical \#1. Once requests are admitted, the operator allocates the required shared and dedicated physical/virtual network functions (P/V NFs), resources, radio access technologies (RATs) settings, and per slice tailored user/control-plane splits to every US-NSI according to its NST. These multiple US-NSIs' requests may or may not be granted concurrently. Based on this, a dynamic, automated, and synchronized strategy of interactions is anticipated in the northbound to fulfill the technical requirements of each of the US-NSIs.

The interactions between operator and tenant (e.g., dynamic and automated negotiation over SLA, slice request, and network resources) are accomplished by allowing the service capability exposure function (SCEF) to securely expose required services and capabilities to every US-NSI. The SCEF, specified by the 3GPP in Release 13, is one of the nodes in LTE core network that acts as a mediator between operator and tenant. In 5GC, the network exposure function (NEF) was introduced to evolve the capabilities of SCEF with special emphasis on network slicing. The SCEF/NEF abstracts services from the 3GPP network's interfaces and protocols to a tenant (see Fig. \ref{fig:US-NSI.pdf}). Moreover, it allows a tenant to configure, monitor, control and change certain policies, QoS, and status of the user equipment (UE) attached to a US-NSI. These unique capabilities of SCEF/NEF provide operators the opportunity to fulfill the requirements of innovative use cases and more flexibly offer services that they could not support prior to its deployment.  

The operator offers these capabilities to a tenant over a set of dedicated northbound application programming interfaces  (APIs or T8), which is standardized by 3GPP in \cite{3GPPTS23682}. By means of API, the tenant can configure resources, communication services, and NFs tailored to every US-NSI. In the proposed architecture, an independent set of APIs is considered to every use case. Fig. \ref{fig:US-NSI.pdf} depicts that for $n$ use cases of Vertical \#1, $n$ sets of independent APIs are provided. These APIs connect the SCEF/NEF and the service capability server/application server (SCS/AS). 

The SCS/AS acts as a gateway between operator and tenant. It may or may not reside at the operator's trust domain. In this architecture, it is assumed that the SCS/AS is provided by the tenant. Fig. \ref{fig:US-NSI.pdf} illustrates that every use case has been provided with its own SCS/AS. The SCEF/NEF provides the SCS/AS access to a rich set of information related to US-NSI. The SCS/AS exposes this information to obtain access to a significant amount of insight into UE that its corresponding US-NSI supports, such as users density, status and location of UE, connectivity, reachability, failure in the communication link, retainability, etc. This brings a number of advantages including energy efficiency, spectrum efficiency, simplicity in UE management, etc. 

\begin{figure*}[!h]
    \centering
    \includegraphics[scale=0.35]{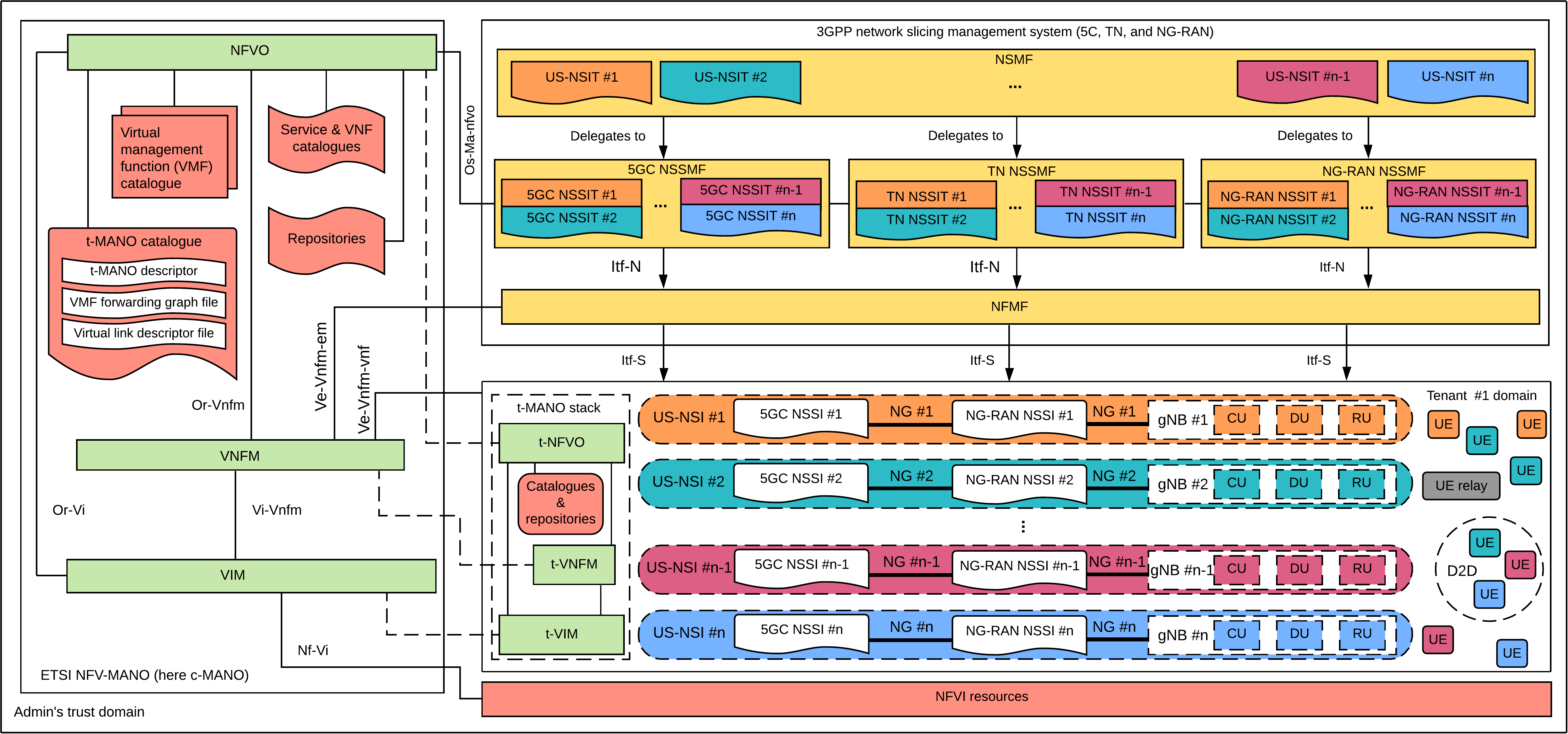}
    \caption{The management and orchestration of the US-NSIs over 3GPP/NFV-based slicing framework.}
    \label{fig:access.pdf}
\end{figure*}

The 5GC does also consist of network slice broker (NSB), which enables the tenant to dynamically request and lease resources form the operator, based on US-NSIs' service requirements that are translated into technical requirements in their NSTs \cite{7514161}. The NSB is located in the operator's network management system (NMS) and connected with the SCEF/NEF through APIs to tenants (see Fig. \ref{fig:US-NSI.pdf}). It dynamically allocates network resources, admission control, and NG-RAN scheduler configuration to every US-NSI.

Moreover, there is a number of new components introduced by 3GPP in both control and user planes for an efficient NFV, dynamic resource allocation, and flexible multi-tenant deployment (see Fig. \ref{fig:US-NSI.pdf}). On the control plane, the access and mobility function (AMF) supports authentication and registration of UE, the session management function (SMF) manages multi sessions contexts, and the policy control function (PCF) provides policy rules to control plane functions \cite{3GPPTS29513V1500}. On the user plane, user plane function (UPF) forwards and routes packets between UEs of US-NSI and data network through NG-RAN \cite{3GPPTS29513V1500}. These components are interconnected using 3GPP defined interfaces (see Fig. \ref{fig:US-NSI.pdf}).

The NG-RAN components are also virtualized in terms of application, cloud, spectrum, and cooperation. The main objectives of virtualization in the NG-RAN are to map and implement US-NSIs in an efficient manner from the perspectives of energy, cost, radio resources, spectrum, and security; while guaranteeing the agreed SLA for each US-NSI. The NG-RAN (specifically the gNB) interacts over N2 (or next-generation [NG]) interface with the 5GC (see Fig. \ref{fig:US-NSI.pdf}) in order to provide required NFs and resources to each US-NSI \cite{Ferrus2019}. The gNBs are interconnected over Xn interfaces, and deliver full NG-RAN functionality to interact with the UEs using the NR interface. 

\subsection{The Management and Orchestration of the US-NSIs} \label{subsection:Instantation}
Tenant \#1 requests communication service management function (CSMF, beyond the scope of this article) for the required communication services of its vertical. The CSMF translates these requirements to US-NSIs' requirements and delivers them to the network slice management function (NSMF). The NSMF either selects predefined NSTs from NSC or creates new NSTs for requested US-NSIs. The NST of the US-NSI (i.e., US-NSIT) makes the US-NSI network-wide available and identifies -- using vertical/use-case identification information (see the upper-right side of Fig. \ref{fig:US-NSI.pdf}) -- it uniquely in the tenant’s domain. The NSMF is also responsible for the orchestration and LCM of $n$ US-NSIs, through $n$ US-NSITs, using certain self-organizing network (SON) algorithms (see Fig. \ref{fig:access.pdf}).

Every US-NSI is composed of 5GC network slice subnet instance (5GC NSSI) and NG-RAN NSSI, which are connected over TN NSSI. They are instantiated based on their corresponding templates, i.e., 5GC NSSIT, NG-RAN NSSIT, and TN NSSIT, respectively, which are derived from their associated US-NSIT. The 5GC NSSI is a set of shared/dedicated VNFs/PNFs, which provides a particular 5GC behavior. The NG-RAN NSSI is a set of configured gNBs (including PNFs and VNFs), which provides a particular NG-RAN behavior. The TN NSSI (illustrated by NG interface in Fig. \ref{fig:access.pdf}) is a set of networking and computing resources, which interconnects the 5GC and NG-RAN subnets. 

To manage fault, configuration, accounting, performance, and security (FCAPS), and life-cycles of three aforementioned subnets, the 3GPP defines network slice subnet management function (NSSMF). Every subnet has its own NSSMF, which reports to NSMF using management interfaces. The NSSMF maps the NSSI's requirements to communication, computation, and storage resources in 5GC, TN, and NG-RAN. Each NSSMF is connected to network function management function (NFMF) over Itf-N interface. Each type of component in each of the subnets has its own NFMF. For example, the gNB has three components, namely the centralized unit (CU), distributed unit (DU), and radio unit (RU). Each component of each gNB has its own NFMF in each NG-RAN NSSI and is responsible for the FCAPS of that particular component. The NFMF and the component under its control are interconnected over the Itf-S interface. To avoid duplicated components and redundant information, we put one NFMF in Fig. \ref{fig:access.pdf}. 

\begin{figure*}[!h]
    \centering
    \includegraphics[scale=0.5]{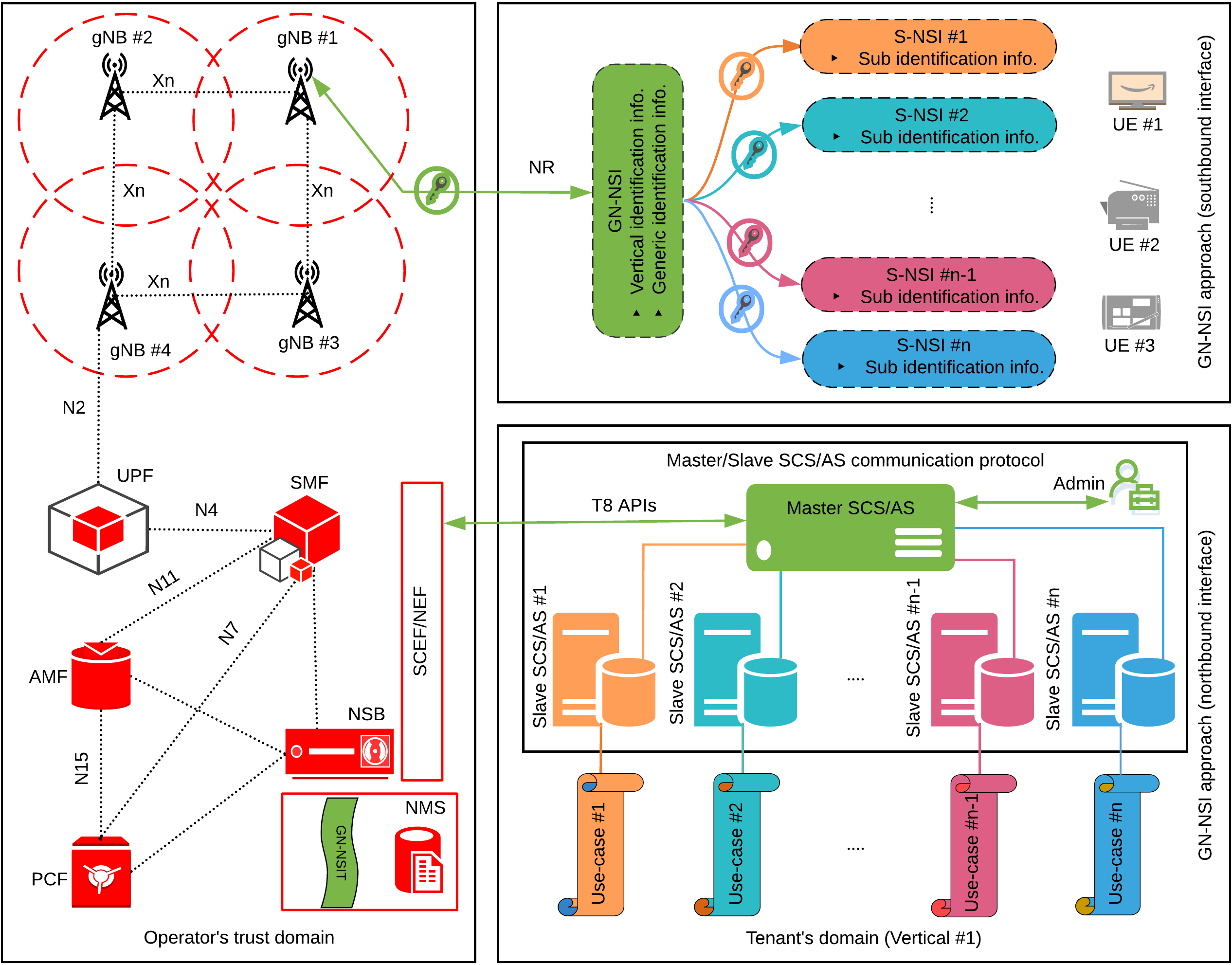}
    \caption{The architectural framework of the Sub Network Slicing.}
    \label{fig:GN-NSI.pdf}
\end{figure*} 

To effectively manage and orchestrate P/V-NFs and their resources, the 3GPP network slicing management system is not enough, because the LCM of the VNFs and the components implementing them are in the scope of the ETSI NFV management and orchestration (MANO) \cite{8713808}. The ETSI and 3GPP have thus jointly proposed a unified framework for the E2E management and orchestration of the 5G network slicing system illustrated in Fig. \ref{fig:access.pdf}. The NFV-MANO (hereinafter MANO) is composed of NFV orchestrator (NFVO), VNF manager (VNFM), and virtualized infrastructure manager (VIM) \cite{8713808}. These three functional blocks are interconnected over standard interfaces, which are shown in Fig. \ref{fig:access.pdf}. Moreover, they are integrated with 3GPP network slicing management entities over standard interfaces for effectively managing and orchestrating the US-NSIs. Further insights into the functional blocks and interfaces of MANO and their tentative integration with 3GPP management entities are out of the scope of this article. The interested readers are advised to refer to the relevant publications.

The current design of MANO provides a centralized management approach, which involves a discrete number of challenges including performance, management, monitoring and processing overload on central (c)-MANO, etc. To overcome these challenges and provide the tenant with the required autonomy for managing and orchestrating the US-NSIs and their corresponding resources, services, and policies, we propose the tenant (t)-MANO in the tenant's domain \cite{8713808}. In this concept, a customized t-MANO instance is abstracted from a c-MANO to a tenant through negotiation and enforcement of management level agreement (MLA), which provides a significant level of control and capabilities to the tenant on requested US-NSIs. This enables the tenant to exercise MANO's functions over US-NSIs with minimum reliance on the c-MANO. However, the c-MANO has full administrative rights and controls on the t-MANO stack, such as monitoring MLA compliance and provide services, features, and capabilities.

The US-NSIs can be offered to the UEs from the same or different gNB(s)\cite{Ferrus2019}. 3GPP also specifies that a UE can be served by at most eight NSIs (i.e., US-NSIs) concurrently over a single NG-RAN, sharing a common AMF in all US-NSIs and an individual SMF to every US-NSI \cite{3GPPTS23501V1540} in the following two scenarios (see the bottom-right side of Fig. \ref{fig:access.pdf}): 

\begin{itemize}
\item \textbf{UEs served by a dedicated US-NSI:} are the UEs of the same service category that are served by their corresponding US-NSI (differentiated by the same color in Fig. \ref{fig:access.pdf}) over their respective NR interfaces throughout an agreed business duration. Such UEs can also act as a relay to facilitate device to device (D2D) communication with other UEs of the same service category.

\item \textbf{UEs served by multiple US-NSIs:} are the UEs that are served by multiple US-NSIs (maximum eight) of different service categories in both intra-frequency and inter-frequency modes, concurrently. Each US-NSI is offered to such a UE through an individual NR interface. These UEs can also facilitate D2D communication with other UEs of their service categories. Due to space limitations, these types of UEs are not illustrated in Fig. \ref{fig:access.pdf}.
\end{itemize}

\begin{figure*}[!h]
    \centering
    \includegraphics[scale=0.23]{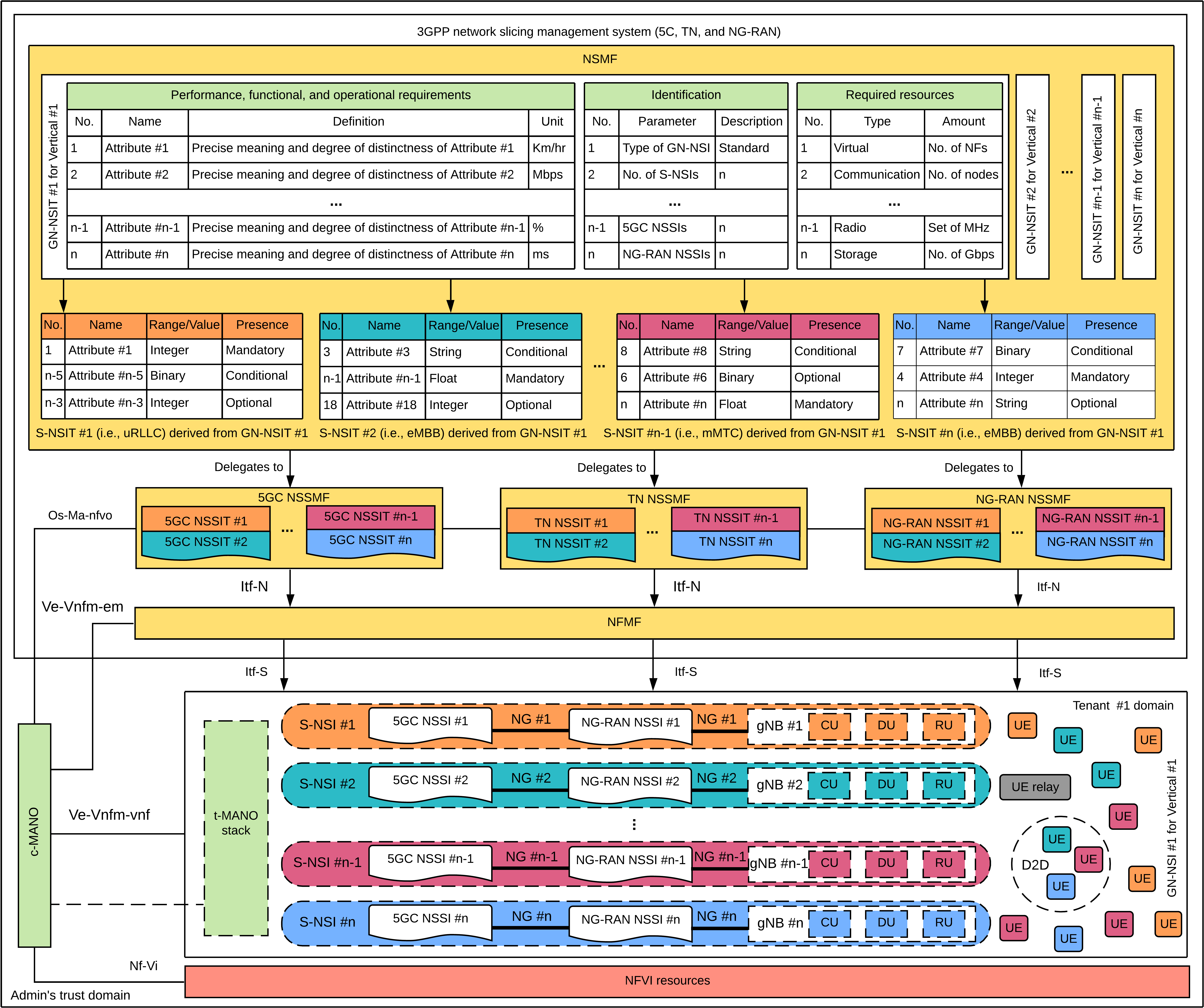}
    \caption{The management and orchestration of S-NSIs in a logical cluster of GN-NSI over 3GPP/NFV-based slicing framework.}
    \label{fig:GN-NSI001.pdf}
\end{figure*}  

\section{Sub Network Slicing} \label{sec:GenericNetworkSlice}
The Sub Network Slicing is an alternative approach to providing network slicing solution to multiple use cases of a vertical. In this concept, a number of NSIs, called the S-NSI (or Micro), are designed to fulfill service requirements of a vertical that consists of multiple use cases. Every S-NSI is dedicated to a use case. These S-NSIs are then logically clustered or bundled (see Fig. \ref{fig:GN-NSI.pdf}) and subsequently provided as a single business product to the tenant. This family (or group) of S-NSIs is called the GN-NSI. The GN-NSI is a per-vertical (or Macro) NSI, which is designed for a specific type of vertical. It does not necessarily convey the meaning of one NSI per vertical, but rather, it is a logical cluster (or a set) of $n$ S-NSIs. 
There are two types of GN-NSIs: 
\begin{itemize}
\item \textbf{Standard GN-NSI:} is a GN-NSI, which is by default defined, designed, and configured with all of its S-NSIs through joint efforts between operators, vendors, verticals, regulators, and SDOs, and its NST is already inserted into the NSC. When the tenant requests a GN-NSI, the type of its vertical has to be identified, its corresponding NST has to be selected from the NSC, and the GN-NSI must be subsequently offered. This approach makes the whole business process, more specifically the preparation phase of the LCM of a GN-NSI, simpler and easier. Both sides can reach a win-win agreement in a short time.
\item \textbf{Non-standard GN-NSI:} is a GN-NSI, which is not available in the NSC or is not yet standardized by the five aforementioned parties. In this case, the operator and tenant privately negotiate to define the attributes, design its NST, and subsequently insert it into the NSC. Both parties equally participate in the configuration of the requirements, the allocation of resources and NFs, and the instantiation of S-NSIs. In this approach, the tenant has more influence over the configuration of the GN-NSI and its corresponding S-NSIs as well as the allocation of required NFs and resources during the whole business period.
\end{itemize}

\subsection{The Architectural Framework of the Sub Network Slicing}
In Fig. \ref{fig:GN-NSI.pdf}, Vertical \#1 (exemplified in the previous section) is assumed again. Each use case is provisioned with an independent S-NSI (i.e., S-NSI \#1 -- S-NSI \#n, respectively). These S-NSIs are then clustered in the form of GN-NSI \#1. Tenant \#1 requests a GN-NSI in the northbound interface (see the bottom-right side of Fig. \ref{fig:GN-NSI.pdf}). The GN-NSI's request coming through a set of APIs indicates its requested NST. The NSB is in charge of granting or denying the GN-NSI's request, and dynamically allocating of resources to the tenant based on a pre-defined SLA. Once the request is granted, the operator provides the desired GN-NSI back through its own set of APIs from SCEF/NEF to SCS/AS.

In order for the tenant to interact with the operator and manage its S-NSIs, we propose the Master/Slave (which is also called the Primary/Secondary) communication protocol in the northbound interface. In this architecture, a single Master-SCS/AS and multi Slave-SCS/ASs (i.e., Slave-SCS/AS \#1 -- Slave-SCS/AS \#n), controlled and managed by a system administrator, are taken into consideration (see Fig. \ref{fig:GN-NSI.pdf}). Both Master/Slave-SCS/ASs must have their own application systems. The Master-SCS/AS requests the GN-NSI from the operator and controls $n$ Slave-SCS/ASs. Each Slave-SCS/AS is customized to a use case and runs a single S-NSI. Once the tenant's request for the GN-NSI is admitted, and the Master-SCS/AS and the Slave-SCS/AS relationship is established, the direction of the S-NSI's request, resource allocation, and control is always from Master to Slave(s).

Through the southbound interface (see the upper-right side of Fig. \ref{fig:GN-NSI.pdf}), the tenant accesses a GN-NSI, containing S-NSIs, over their corresponding NR interfaces. The GN-NSI and its associated S-NSIs can be offered from a single or multiple gNB(s). Some radio processing functions and resources are shared among S-NSIs, whereas some are dedicated to S-NSI \cite{8703477}. The configuration and allocation of resources and NFs of S-NSIs in a logical cluster of GN-NSI in the NG-RAN is a challenging issue. Further research is thus required to analyze the feasibility and efficiency of Sub Network Slicing in the NG-RAN.

The UEs may or may not access the S-NSIs concurrently. For example, if a vehicle connected to a V2X GN-NSI is parked in a parking area. Two S-NSIs (i.e., related to infotainment applications and autonomous driving) are presumably disconnected, whereas the third S-NSI (i.e., related to vehicle management) may be operating. However, if a vehicle is in the driving mode, then all three S-NSIs, in a cluster of GN-NSI, are simultaneously provided to the vehicle. The UE can not access more than eight S-NSIs concurrently in the NG-RAN \cite{3GPPTS38300}. Therefore, a GN-NSI may consist of up to eight S-NSIs. The UE access scenarios to the S-NSIs of a GN-NSI (see Fig. \ref{fig:GN-NSI001.pdf}) are also the same as access to the US-NSI discussed in section \ref{subsection:Instantation}.

\subsection{The Management and Orchestration of the S-NSIs}
The CSMF receives a list of service requirements from Tenant \#1, translates it to the GN-NSI's requirements, and delivers it to the NSMF. The NSMF either selects a predefined NST from NSC or creates a new NST for requested GN-NSI. The NST of a GN-NSI is called the GN-NSI template (GN-NSIT). The NSMF is also responsible for the LCM of $n$ GN-NSIs through their corresponding $n$ GN-NSITs (see Fig. \ref{fig:GN-NSI001.pdf}).

The GN-NSIT is composed of functional, performance, and operational attributes (each attribute with its own definition and unit of measurement), required resources information, and identification information. The two sets of information are in compliance with the 3GPP definition of NST. However, to further identify the S-NSIs and their corresponding components, we propose an identification set of information in the GN-NSIT -- vertical/generic on the GN-NSI level and Sub on the S-NSIs level. These three sets of information, which define a GN-NSI from technical and business perspectives, are subsequently used to obtain a group of S-NSIs, according to the number of the use cases of a vertical, (see Fig. \ref{fig:GN-NSI001.pdf}). 

Each S-NSI has also an NST with industry-accepted characteristics, which is called the S-NSI template (S-NSIT). The attributes that define an S-NSIT are taken from GN-NSIT (see Fig. \ref{fig:GN-NSI001.pdf}). The S-NSIT only includes necessary information related to the range/value (i.e., integer, binary, string, float, etc.) and presence (i.e., mandatory, optional, and conditional) of attributes required by the S-NSI. Once the selection/design of GN-NSIT and its associated S-NSITs are completed. The NSMF is responsible for FCAPS, management, and orchestration throughout their life-cycles. The FCAPS, LCM, and templates of the components of the S-NSI are the same as those of US-NSI. The t-MANO has also introduced here to effectively manage and orchestrate the virtualized part of the S-NSIs in the tenant's domain. 

\section{Conclusions and future outlook} \label{sec:conclusion}
In this article, we have addressed the solution gap in terms of the flexible and dynamic provisioning and efficient management of per-vertical per-use-case NSIs tailored to the respective use case service and resource requirements in order to ensure its functional and operational integrity in isolation from other use cases of a single vertical. In this context, we have presented two solution concepts, namely the Use-case Specific Network Slicing and the Sub Network Slicing, and their architectural frameworks, while placing a special emphasis on their management and orchestration in 5G mobile networks. It is important to mention that both concepts and their corresponding architectures are in compliance with the definition of network slicing, as defined by the NGMN Alliance and the 3GPP. Due to this, it is planned to introduce these two solution concepts in the relevant SDOs, such as in the relevant technical specification groups of 3GPP and ETSI NFV in order to address the solution gap to this important requirement.

In the future, we are interested to investigate the qualitative and quantitative comparison of a number of technical (such as resource allocation, isolation, security, admission control, etc.) and business (such as LCM, capital and operational expenditure, SLA, etc.) related parameters of both concepts in order to figure out their potential deployment implications on 5G and beyond mobile networks. The comparative study of both approaches will give interested readers further insights into the applicability and feasibility of the US-NSI and the GN-NSI architectural frameworks in terms of efficiency, complexity, and flexibility. There is no doubt that the implementation of both concepts is also going to raise a number of technical challenges for the UE, the NG-RAN, and the 5GC. Exploring these research problems in the future is thus essential with the purpose of motivating new advances and potential solutions related to the US-NSIs and the GN-NSIs.

\section*{Acknowledgments}
The authors are deeply grateful to the anonymous reviewers for their valuable remarks and insightful comments, which significantly contributed to the quality of this article, and to the Editor-in-Chief for coordinating the peer-review process. This work has been partially supported by the H2020-MSCA-ITN-2015 project 5G AuRA (Grant 675806) and the 5G-CARMEN project (Grant 825012).

\bibliography{ref/mypaper01.bib} 
\bibliographystyle{ieeetr}

\vskip 0pt plus -1fil

\section*{Author Biographies}
\begin{IEEEbiography}[{\includegraphics[width=1in,height=1.25in,clip,keepaspectratio]{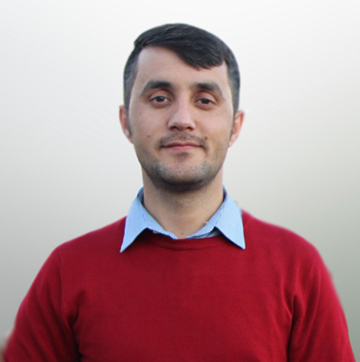}}]{Mohammad Asif Habibi} (asif@eit.uni-kl.de) received his B.Sc. degree in Telecommunication Engineering from Kabul University-Afghanistan in 2011, and his M.Sc. degree in Systems Engineering and Informatics from Czech University of Life Sciences-The Czech Republic in 2016. Since January 2017, he has been working as Research Fellow and Ph.D. Candidate at Technische Universit\"at Kaiserslautern, Germany.
\end{IEEEbiography} 
\vskip 0pt plus -1fil

\begin{IEEEbiography}[{\includegraphics[width=1in,height=1.25in,clip,keepaspectratio]{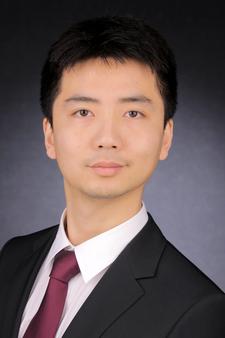}}]{Bin Han} (binhan@eit.uni-kl.de) received in 2009 his B.E. degree from Shanghai Jiao Tong University, M.Sc. in 2012 from Technische Universit\"at Darmstadt, and in 2016 the Dr.-Ing. degree from Karlsruhe Institute of Technology. Since July 2016 he has been with Technische Universit\"at  Kaiserslautern, researching in the broad area of wireless networks and signal processing.
\end{IEEEbiography} 
\vskip 0pt plus -1fil

\begin{IEEEbiography}[{\includegraphics[width=1in,height=1.25in,clip,keepaspectratio]{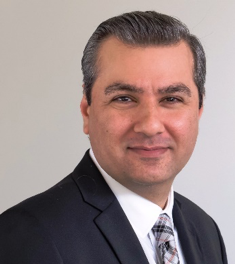}}]{Faqir Zarrar Yousaf} (zarrar.yousaf@neclab.eu) is a senior researcher at NEC Laboratories Europe, Germany. He completed his Ph.D. at TU Dortmund, Germany. His current research interest is in NFV/SDN in the context of 5G networks. He is also a delegate to the ETSI NFV standards organization, where he is a Rapporteur for four work items and has contributed to several standards. He has one granted patent and 16 filed patents, and his research work has been widely published.
\end{IEEEbiography} 
\vskip 0pt plus -1fil

\begin{IEEEbiography}[{\includegraphics[width=1in,height=1.25in,clip,keepaspectratio]{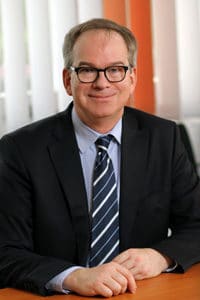}}]{Hans D. Schotten} (schotten@eit.uni-kl.de) received the Diploma and Ph.D. degrees in electrical engineering from the Aachen University of Technology RWTH, Germany in 1990 and 1997, respectively. Since August 2007, he has been a full professor and head of the Institute of Wireless Communication and Navigation at Technische Universit\"at Kaiserslautern. Since 2012, he has also been Scientific Director at the German Research Center for Artificial Intelligence, heading the ``Intelligent Networks'' department.
\end{IEEEbiography}
\end{document}